\documentclass{Interspeech2024}

\interspeechcameraready

\title{Accent Conversion with Articulatory Representations}

\name[affiliation={1}]{Yashish M.}{Siriwardena}
\name[affiliation={2}]{Nathan}{Swedlow}
\name[affiliation={2}]{Audrey}{Howard}
\name[affiliation={2}]{Evan}{Gitterman}
\name[affiliation={2}]{Dan}{Darcy}
\name[affiliation={1}]{Carol}{Espy-Wilson}
\name[affiliation={2}]{Andrea}{Fanelli}

\address{
  $^1$University of Maryland College Park, MD, USA\\
  $^2$Dolby Laboratories, CA, USA}
\email{yashish@terpmail.umd.edu, Nathan.Swedlow@dolby.com, audrey.howard@dolby.com, evan.gitterman@dolby.com, Dan.Darcy@dolby.com, espy@umd.edu, Andrea.Fanelli@dolby.com}


\begin{document}

\maketitle

\begin{abstract}


Conversion of non-native accented speech to native (American) English has a wide range of applications such as improving intelligibility of non-native speech. Previous work on this domain has used phonetic posteriograms as the target speech representation to train an acoustic model which is then used to extract a compact representation of input speech for accent conversion. In this work, we introduce the idea of using an effective articulatory speech representation, extracted from an acoustic-to-articulatory speech inversion system, to improve the acoustic model used in accent conversion. The idea to incorporate articulatory representations originates from their ability to well characterize accents in speech. To incorporate articulatory representations with conventional phonetic posteriograms, a multi-task learning based acoustic model is proposed. Objective and subjective evaluations show that the use of articulatory representations can improve the effectiveness of accent conversion.

\end{abstract}

\vspace*{-6pt}
\section{Introduction}
\label{sec:intro}

Foreign Accent Conversion (FAC) is the process of transforming non-native English speech to match the accent or pronunciation pattern of a native English speaker, while retaining the speaker identity \cite{zhao_journel}. The defining step of every FAC pipeline is the extraction of meaningful speaker-independent speech embeddings, that disentangle every other speaker-specific vocal element from the accent. In most of the FAC pipelines, the Acoustic Model (AM) is the model pre-trained on native speech (eg. Librispeech \cite{Panayotov2015LibrispeechAA}) to estimate such speech embeddings. Different versions of phonetic posteriograms (PPGs) have been previously used as speech embeddings for accent conversion \cite{zhao_accent_ppgs, zhao_accent_ppg_pairing, zhao_journel}, capturing how individual sounds are produced in speech over time. However, the higher dimensionality of PPGs, combined with their sparse nature, have led to the idea of using compact bottle neck features (BNFs) -- features normally extracted from a layer of the pre-trained acoustic model -- for the downstream accent conversion \cite{zhao_journel, quamer22_interspeech, quamer23_interspeech}. 


Techniques used for accent conversion can be broadly categorized into acoustic and articulatory methods \cite{zhao_journel}. Articulatory-based methods use speaker's articulatory trajectories -- absolute x, y, z coordinates of the pallets placed on articulators, (e.g. lips, tongue, jaw) -- to train a speaker-specific articulatory synthesizer. Here the synthesizer is trained to learn a mapping from the L2 speaker's articulatory data to an acoustic feature (e.g. Mel Cepstra). To perform accent conversion, the trained articulatory synthesizer is then provided with articulatory trajectories from a native speaker \cite{articulatory_sythesis_1, ARYAL2016260}. Disentangling accent from voice identity in the articulatory domain is intuitive and effective, but it is impractical due to the challenges in collecting actual ground-truth articulatory data, which is expensive and requires specialized equipment \cite{zhao_journel}. On the other hand, acoustic methods are more applicable since they only require speech in acoustic form, but can be less effective in decoupling the accent from the other important vocal traits \cite{zhao_journel}. 

In this paper, we explore the possibility of using relative articulatory measures (Browman et al. \cite{Browman1992}) in place of the absolute articulatory trajectories previously used for accent conversion. These relative articulatory measures, which are referred to as vocal tract variables (TVs), are more speaker-independent in nature when compared to the absolute articulatory trajectories. Because of that, they can be effective in developing zero-shot accent conversion systems without training any speaker-specific speech synthesizer. To circumvent the need of obtaining ground-truth articulatory data (TVs), we use a state-of-the-art acoustic-to-articulatory speech inversion system \cite{attia_yashish_2023improving}, originally trained with ground-truth articulatory data from the Wisconsin X-ray microbeam dataset \cite{Westbury1994b}. One of the key contributions of this work is the development of a new acoustic model to extract speech embeddings for the accent conversion pipeline. In our model, the TVs extracted from the speech inversion system are used along with PPGs in a multi-task learning framework, to address whether utilizing PPGs or TVs alone, or a combination of both, can enhance the downstream task of accent conversion. 

\vspace*{-6pt}
\section{Related Work}
\label{sec:related_work}

Foreign accent conversion with articulatory representations have been previously explored in \cite{ARYAL2016260, aryal-2016-interspeech, felps2012taslp}, where speaker-specific articulatory synthesizers are first trained with ground-truth Electro Magnetic Articulograpy (EMA) data. As discussed in \cite{aryal-2016-interspeech}, these articulatory based synthesizers have struggled in synthesizing speech with good acoustic quality compared to that of acoustic feature based synthesizers (eg. MFCCs). The incomplete representation of the vocal tract, compounded with the difficulty in collecting articulatory data, has led the speech community to resort to purely acoustic based methods to perform accent conversion. However, recent work with vocal tract variables, an improved articulatory representation, has shown that they can be effectively used for high quality speech synthesis \cite{wu22i_interspeech, siriwardena23_interspeech}. This also suggests that effective articulatory representations with improved deep learning algorithms can enable accent conversion with articulatory representations.   

Apart from categorizing the accent conversion pipelines to articulatory and acoustic based methods, they can also be discussed based on the use (or not) of native reference speech. Accent conversion systems developed over the years have used reference native speech to extract the native pronunciation pattern to perform accent conversion \cite{DING2022101302, zhao_accent_ppgs, zhao_accent_ppg_pairing, quamer23_interspeech}. Since the applications of this approach are limited, recent work has mostly explored ways to perform reference-free accent conversion \cite{Liu_reference_free, zhao_journel, Wang_reference_free, quamer22_interspeech, Jin_reference_free}. In a recent work by Quamer et al., \cite{quamer23_interspeech}, an accent conversion pipeline with a transformer based seq2seq model was introduced. The accent conversion pipeline here uses a native reference and involves an acoustic model trained to estimate senone-PPGs. The work there is rather exploratory and shows that segmental and prosodic cues of non-native speech can be disentangled. We adopted the training paradigm and the seq2seq synthesizer architecture from this work and modified it to perform accent conversion with a newly designed acoustic model to incorporate articulatory representations.

\vspace*{-6pt}
\section{Method}
\label{sec:method}

\subsection{Articulatory representations from Speech Inversion}
\label{ssec:tv_extraction}
\vspace*{-3pt}

Acoustic-to-articulatory speech inversion (SI) aims to infer articulatory dynamics from spoken sounds \cite{Sivaraman_ASA}. While efforts to interpret articulatory movements from continuous speech signals have a long history \cite{Papcun1992}, they have typically been limited to tracking specific parts of the vocal tract, like the upper and lower lips, tongue tip, and velum closure. However, it's essential not only to understand the primary effects of individual vocal tract movements, but also to grasp how these articulators interact. For instance, articulators such as the lips and jaw often cooperate to achieve specific vocal tract shapes \cite{Browman1992}. Consequently, general SI systems prioritize understanding vocal tract constriction, estimating the degree and position of functional tract variables (TVs; from Articulatory Phonology in \cite{Browman1992}), rather than solely focusing on individual articulator movement. During SI, acoustic features extracted from speech signals are used to predict these tract variables. This process involves learning an inverse mapping by training on a dataset containing matched acoustic and directly observed articulatory data.

\begin{figure}[th]
  \centering
  \includegraphics[width=\linewidth]{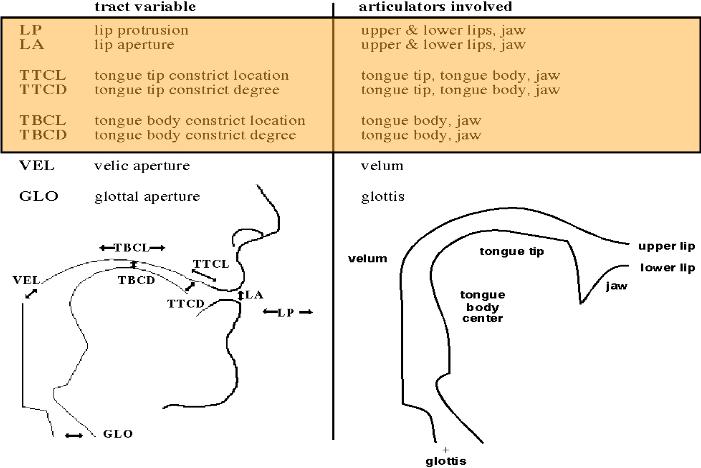}
  \caption{Vocal tract variables and related articulators. The color shaded TVs are extracted from a speech inversion system}
  \label{fig:tv_overview}
  \vspace*{-8pt}
\end{figure}

To train the acoustic model discussed in section \ref{ssec:am_model}, TVs extracted from the SI system in \cite{attia_yashish_2023improving} were used as the ground-truth. The SI system trained and evaluated in speaker-independent fashion estimates six functional TVs. Figure \ref{fig:tv_overview} shows how each of the 6 TVs can be visualized with respect to a vocal tract and the corresponding articulators involved.  

\vspace*{-6pt}
\subsection{Multi-task learning based Acoustic Model}
\label{ssec:am_model}
\vspace*{-3pt}

The proposed multi-task learning based acoustic model (AM) contains two bidirectional long short memory (BiLSTM) layers followed by up-sampling, dropout and fully connected (Linear) layers as shared model layers. A fully connected layer with softmax activation is used to estimate PPG outputs and a fully connected layer with tanh activation is used to estimate TV outputs. Figure \ref{fig:am_model} shows details of the model architecture implemented in PyTorch. The model is optimized using a combined loss ($AM Loss_{combined}$) as defined in equation \ref{eq:am_loss}. A hyperparameter ($\alpha$) is used to weight the two individual losses incurred in estimating PPGs ($PPG_{loss}$) and TVs ($TV_{loss}$). The $PPG_{loss}$ is the cross entropy loss between the estimated and the ground-truth senones. The $TV_{loss}$ is the mean absolute error (MAE) between the estimated and the ground-truth TVs.

\vspace*{-10pt}
\begin{equation}
  \vspace*{-6pt}
  AM Loss_{combined} = \alpha \times TV_{loss} + (1-\alpha) \times PPG_{loss}
  \label{eq:am_loss}
  \vspace*{-4pt}
\end{equation}
\vspace*{-6pt}

Three values of alpha were explored: $\alpha$ = 1, which we will refer to as the `TV only' model, $\alpha$ = 0.4, which we will refer to as the `combined' model and $\alpha$ = 0, which we will refer to as the `PPG only' model. The $\alpha$ for the combined model was chosen by doing a grid search across [0, 0.2, 0.3, 0.4, 0.5, 0.7, 1.0], and checking the average Pearson's product moment correlation (PPMC) scores for TV estimation and RMSE for PPG estimation. $\alpha$ = 0.4 gave the best compromise between the PPMC scores for estimated TVs and loss for estimating PPGs. 

The learning rates to train the models were determined based on a grid search by testing all combinations from [1e-2, 1e-3, 1e-4, 3e-4] that resulted in 1e-4 as the best pick. A similar grid search was done to choose the batch size from [4, 8, 12, 16] and a batch size of 8
gave the best validation loss. The objective function was optimized using the ADAM optimizer with an ‘ExponentialLR’ learning rate scheduler and a decay of 0.5. All models were trained with an early stopping criteria monitoring the validation loss and using a patience of 6 epochs. Once the models are trained, the BNFs are extracted from the final layer of the `shared model layers' as shown in figure \ref{fig:am_model}.

\vspace*{-6pt}
\subsubsection{Dataset and input acoustic features}
\vspace*{-3pt}

\begin{figure}[th]
  \centering
  \includegraphics[width=\linewidth]{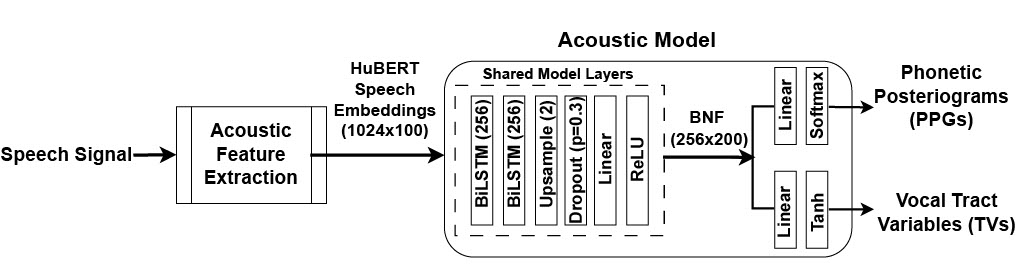}
  \caption{Proposed Multi-task learning based Acoustic model from which PPG only, TV only and Combined BNFs are extracted for accent conversion}
  \vspace*{-8pt}
  \label{fig:am_model}
\end{figure}

Original train, dev and test splits (both -clean and -other) from the LibriSpeech dataset \cite{Panayotov2015LibrispeechAA} was used to train all the acoustic model variants. All the audio files were first segmented to 2 second long segments and the shorter ones were zero padded at the end. As shown in figure \ref{fig:am_model}, the acoustic model takes in Hidden-Unit BERT (HuBERT) \cite{hubert_ppr} speech embeddings extracted from the pre-trained HuBERT-large model as the input speech representation. The HuBERT speech embeddings are sampled at 50 Hz and have a dimensionality of 1024. Tri-phone Phonetic Posteriograms (PPGs) are extracted from a pretrained model \cite{zhao_accent_ppgs} as one of the target speech representations. The extracted PPGs are sampled at 100Hz and have a dimensionality of 5816. As the other target speech representation, TVs are extracted from a pre-trained acoustic-to-articulatory speech inversion system. The TVs are sampled at 100 Hz and contain 6 distinct variables as discussed in section \ref{ssec:tv_extraction}. 

\vspace*{-6pt}

\vspace*{-8pt}
\subsection{Accent Conversion Pipeline}
\label{ssec:acc_pipeline}
\vspace*{-3pt}

\begin{figure}[th]
  \centering
  \includegraphics[width=\linewidth]{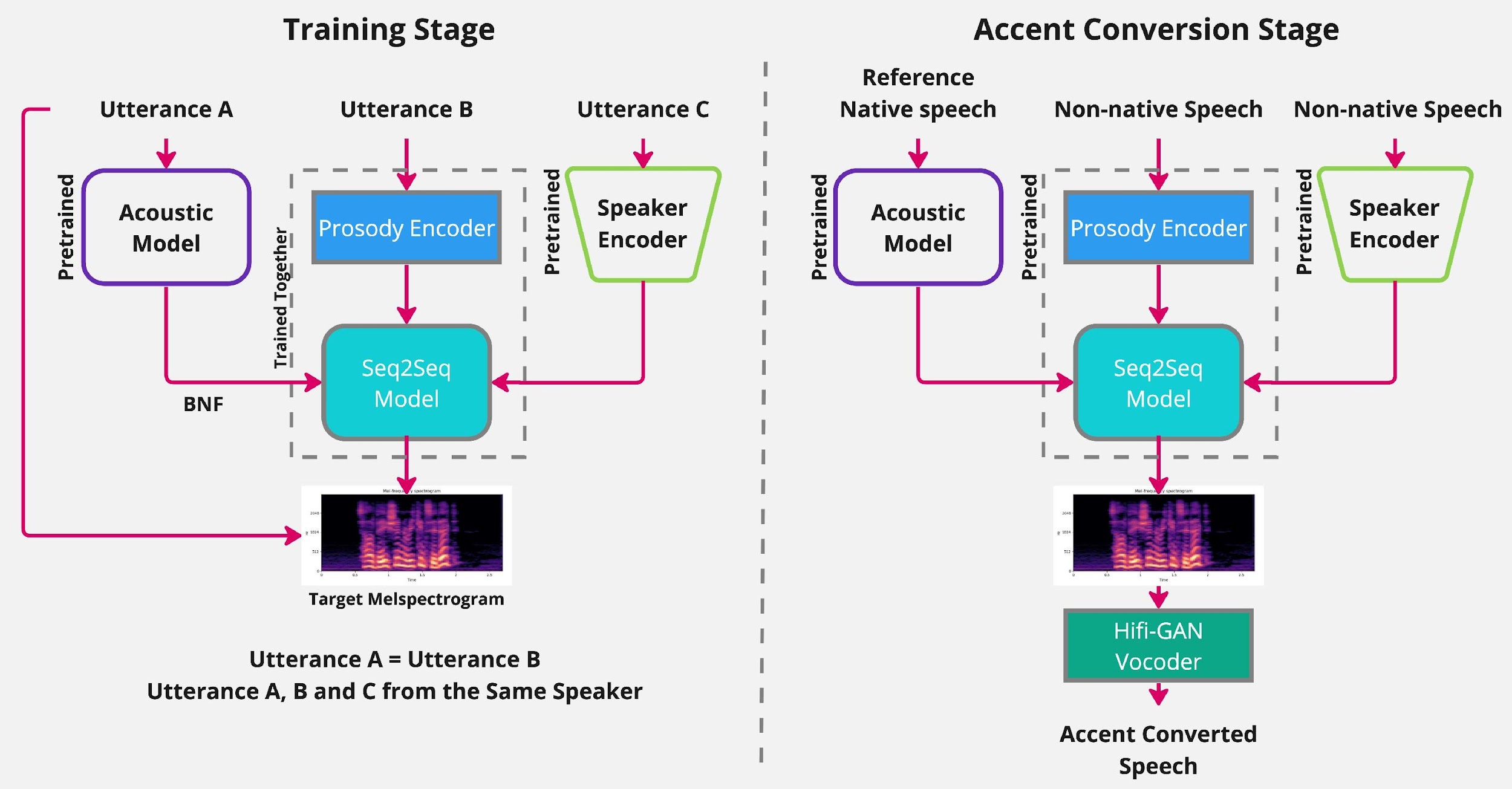}
  \caption{Training and Accent conversion stages of the FAC pipeline}
  \label{fig:fac_architecture}
  \vspace*{-9pt}
\end{figure}

Our experiments were conducted using the ARCTIC \cite{kominek04b_ssw} and L2-ARCTIC \cite{zhao18b_interspeech} datasets. This combined dataset comprises recordings from 28 speakers, each providing 1,132 utterances. Four speakers (NJS, YKWK, TXHC, and ZHAA) were excluded from the training set to serve as unseen speakers during testing. Same train and dev splits as in \cite{quamer23_interspeech} were used. Speaker BDL from the ARCTIC dataset was chosen as the reference native language (L1) speaker for all experiments. For each utterance, we extracted 80-dimensional Mel-spectrograms using a 25ms window and a 10ms shift. To convert Mel-spectrograms into waveforms, we utilized a pre-trained HiFi-GAN vocoder \cite{hifigan_kong}. Figure \ref{fig:fac_architecture} details the foreign accent conversion pipeline used. The model architecture and the training procedure is adopted from the work in \cite{quamer23_interspeech} and comprises of a prosody encoder and a seq2seq synthesizer. 

\vspace*{-6pt}
\vspace*{-2pt}

The acoustic model (AM) transforms an input `utterance A' into a bottleneck feature (BNF) embedding, which encapsulates both the phonetic content and the articulatory constructs of the utterance. The seq2seq model takes in three inputs: BNFs from the AM, a speaker embedding representing the voice characteristics from `utterance C' from a specific speaker, and a prosody embedding from a reference `utterance B'. Utilizing these three sources of information, the seq2seq model tries to reconstruct the original Mel spectrogram. The prosody encoder and seq2seq model are trained synchronously in an unsupervised manner, similarly to an auto-encoder, while the speaker encoder \cite{quamer23_interspeech} and acoustic model are pre-trained beforehand. Throughout training, the acoustic model and prosody encoder are provided with the same utterance from the same speaker, while the speaker encoder receives a different utterance C from the same speaker to generate a speaker embedding. This strategy ensures that the prosody encoder learns a distinct mapping from the speaker encoder, and the seq2seq model avoids inferring prosody solely from the speaker embedding.

Once the prosody encoder and the seq2seq synthesizer are trained, accent conversion is performed as shown in the right panel of Figure \ref{fig:fac_architecture}. Here, a reference speech from a native speaker (matching linguistic content with the L2 speech to be converted) is used to generate the BNFs from the pre-trained acoustic model. A mel-spectrogram extracted from the non-native speech utterance (which needs to be accent converted) is fed to the prosody encoder, and a speaker embedding extracted from the same utterance with the speaker encoder is fed to the seq2seq synthesizer. The synthesized mel-spectrogram as shown in figure \ref{fig:fac_architecture} is then passed through the HiFi-GAN vocoder \cite{hifigan_kong} to generate the accent converted audio waveform. 


\vspace*{-8pt}
\section{Results}
\label{sec:results}
\vspace*{-3pt}

This section summarises the objective and subjective evaluations of the accent conversion pipelines. Synthesized audio samples can be found in the web page\footnote[1]{https://yashish92.github.io/Accent-conversion-TVs/}. 

\vspace*{-8pt}
\subsection{Objective Evaluations}
\label{ssec:quant_analysis}
\vspace*{-3pt}

\subsubsection{t-distributed Stochastic Neighbor Embedding (t-SNE) visualizations}
\vspace*{-4pt}

\begin{figure}[th]
    \centering
    \includegraphics[width=\linewidth]{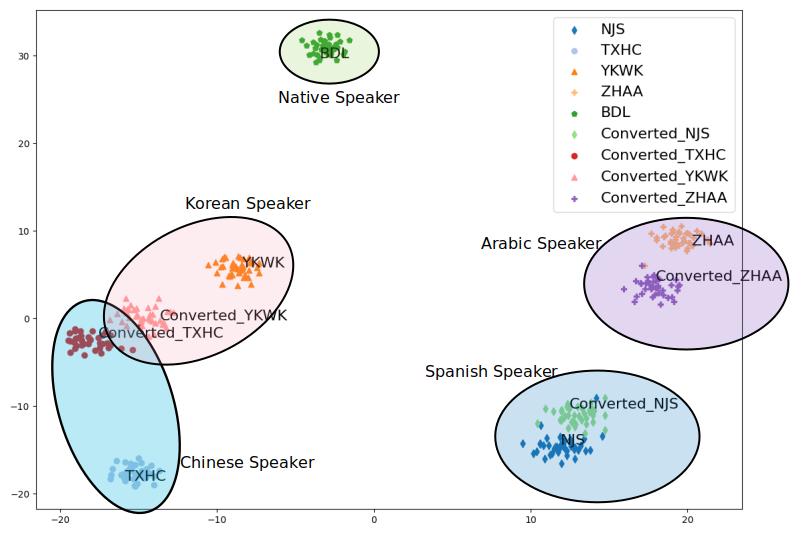}
    \caption{t-SNE visualizations of speaker embeddings from original L2 speech and corresponding accent converted samples from TV only variant}
    \label{fig:tsne_plot}
    \vspace*{-12pt}
\end{figure}

Figure \ref{fig:tsne_plot} shows the t-SNE visualization of the speaker embeddings generated by the speaker encoder for original L2 speech samples and the corresponding accent converted samples of the unseen test set data. Here the visualizations are only provided for the TV only variant since it reported the lowest average distance (9.25 $\pm{5.1}$) between the cluster centroids for accent converted and corresponding original L2 samples. Accent converted speech clustering closer to original speaker’s samples suggests that the TV only model is preserving the speaker's identity noticeably well. It can also be seen that a better speaker identity transfer has happened with Spanish and Arabic speakers (NJS and ZHAA) compared to Korean and Chinese speakers (YKWK and TXHC). 

\vspace*{-8pt}

\vspace*{-8pt}
\subsubsection{Mel Cepstral Distortion (MCD)}

\vspace*{-6pt}
\begin{table}[th]
  \vspace*{-5pt}
  \caption{MCD results from the 3 acoustic model variants. Lower values suggest better synthesis quality}
  \vspace*{-8pt}
  \label{tab:mcd_results}
  \centering
  \begin{tabular}{llll}
    \toprule
    &\textbf{Combined}     &\textbf{TV only}      &\textbf{PPG only} \\
    \midrule
    NJS(Spanish)     &5.8574   &5.8633      &5.7314                          \\
    TXHC (Mandarin)  &6.5491   &6.4276      &6.3474                          \\
    YKWK (Korean)    &6.7301   &6.6984      &6.7331                          \\
    ZHAA (Arabic)    &5.6177   &5.6040      &5.4922                          \\
    \textbf{Average} &6.3719   &6.1614      &\textbf{6.0918}                          \\
    \bottomrule
    \vspace*{-18pt}
  \end{tabular}
\end{table}

\vspace*{-8pt}
\subsubsection{Word Error Rate (WER)}
\vspace*{-10pt}

\begin{table}[th]
  \caption{WER results for the 3 acoustic model variants. Lower scores suggest better recognition by the Whisper-large ASR system \cite{whisper} and hence lesser accentedness}
  \vspace*{-8pt}
  \label{tab:wer_results}
  \centering
  \resizebox{\columnwidth}{!}{
  \begin{tabular}{lllll}
    \toprule
    &\textbf{Combined} &\textbf{TV only}  &\textbf{PPG only}  &\textbf{Original audio}\\
    \midrule
    NJS(Spanish)     &14.70   &14.62      &18.75    &74.70                      \\
    TXHC (Mandarin)  &21.83   &20.67      &25.16    &134.96                      \\
    YKWK (Korean)    &15.57   &15.04      &20.08    &152.73                         \\
    ZHAA (Arabic)    &19.72   &20.99      &23.38    &76.48                       \\
    \textbf{Average} &17.96   &\textbf{17.83}   &21.84    &134.72                        \\
    \bottomrule
    \vspace*{-18pt}
  \end{tabular}}
\end{table}

\vspace*{-10pt}
\subsection{Subjective Analysis}
\label{ssec:subj_analysis}

\begin{table}[th]
  \vspace*{-6pt}
  \caption{Accent Conversion Test ratings (higher values for less perceived foreign accent) and Acoustic Quality Test ratings (higher scores for better quality) for the PPG only, TV only and combined systems compared to L2 and L1 references }
  \vspace*{-6pt}
  \label{tab:test1and2_results}
  \centering
  \resizebox{\columnwidth}{!}{
  \begin{tabular}{llllll}
    \toprule
    &\textbf{Combined} &\textbf{TV only}    &\textbf{PPG Only} &\textbf{Reference L2} &\textbf{Reference L1}\\
    \midrule
    Accent Conversion   &\textbf{7.32 ±0.51}  &7.21 ±0.43 &6.78 ±0.43             &2.07 ±0.37 &8.56 ±0.14 \\
    Voice Quality (MOS) &2.45 ±0.33  &2.59 ±0.36 &\textbf{2.69 ±0.33}            &3.36 ±0.43 &3.46 ±0.18 \\
    \bottomrule
    \vspace*{-26pt}
  \end{tabular}}
\end{table}

We conducted two experiments to evaluate the performance of the TV only, PPG only, and combined FAC systems. Each of these tests investigated a key attribute of interest including accentedness and acoustic quality respectively. Stimulus ordering was randomized and counter balanced across all subjects for both tests. All subjects who participated in these experiments are fluent English speakers and United States citizens. 20 subjects completed each assessment and all subjects were paid to participate. Each test was administered using an internal browser-based testing tool and subjects were instructed to perform each test from a quiet workspace of their choice using headphones for audio playback. 

\vspace*{-5pt}
\subsubsection{Accent Conversion Test}
\vspace*{-3pt}
    
Subjects were tasked with providing a score of accent conversion across 75 audio signals using a nine-point Likert scale, where a score of 9 corresponds to no foreign accent and a score of 1 corresponds to heavy foreign accent. Similar experimental methods have been performed to assess the efficacy of accent conversion by other FAC systems and in these experiments higher scores typically correspond to greater levels of accentedness \cite{Zhao_ASLP,quamer22_interspeech}. Given that our research evaluates the performance of accent conversion systems, we determined that higher subjective scores should represent the effect of accent conversion.  

\begin{figure}[th]
  \centering
  \includegraphics[width=\linewidth]{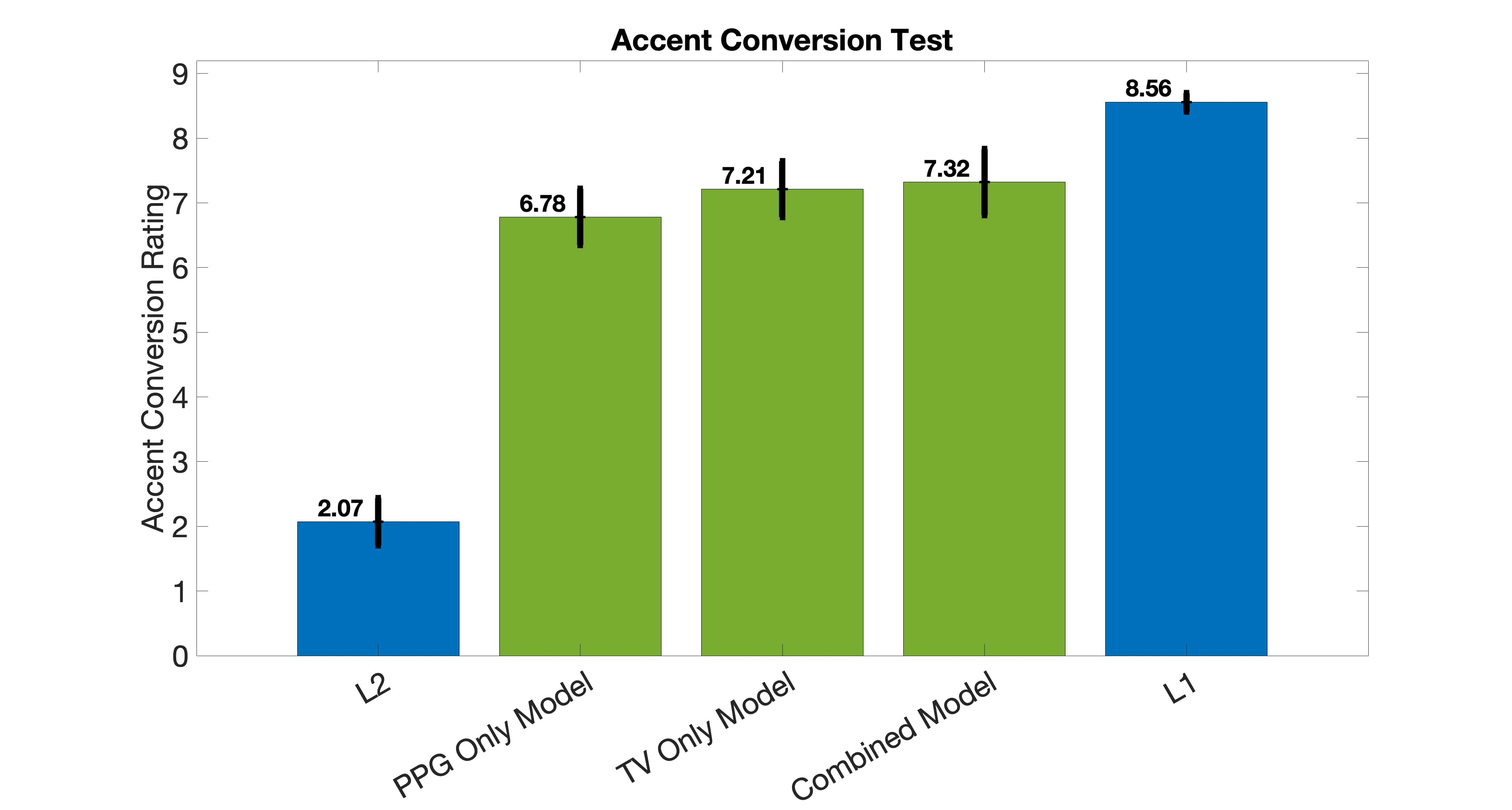}
  \caption{Bars are mean rating for each system. References are shaded blue and test systems are shaded green. Error bars are 95\% confidence interval}
  \label{fig:accentedness}
  \vspace*{-14pt}
\end{figure}

Each subject ranked 15 utterances per test system (PPG Only, TV Only, Combined) resulting in 45 total accent-converted test signals. Subjects also scored 15 L2 and 15 L1 reference utterances. We evaluated the performance of all three test systems relative to the L2 and L1 references using two-sample t-tests with a Bonferroni adjusted significance level ($\alpha = 0.005$). All three FAC systems scored significantly higher than the L2 reference ($p < 0.005$), meaning subjects perceived less foreign accent in the converted speech. Additionally, all three systems scored significantly lower than the L1 reference ($p < 0.005$), meaning greater accentedness was perceived in the test systems relative to the native English speaker reference. The TV only model and combined model scored slightly higher than the PPG only model, however these results were not statistically significant. Table \ref{tab:test1and2_results} and figure \ref{fig:accentedness} shows the results.   

\vspace*{-6pt}
\subsubsection{Acoustic Quality Test}
\vspace*{-3pt}

\begin{figure}[th]
  \centering
  \includegraphics[width=\linewidth]{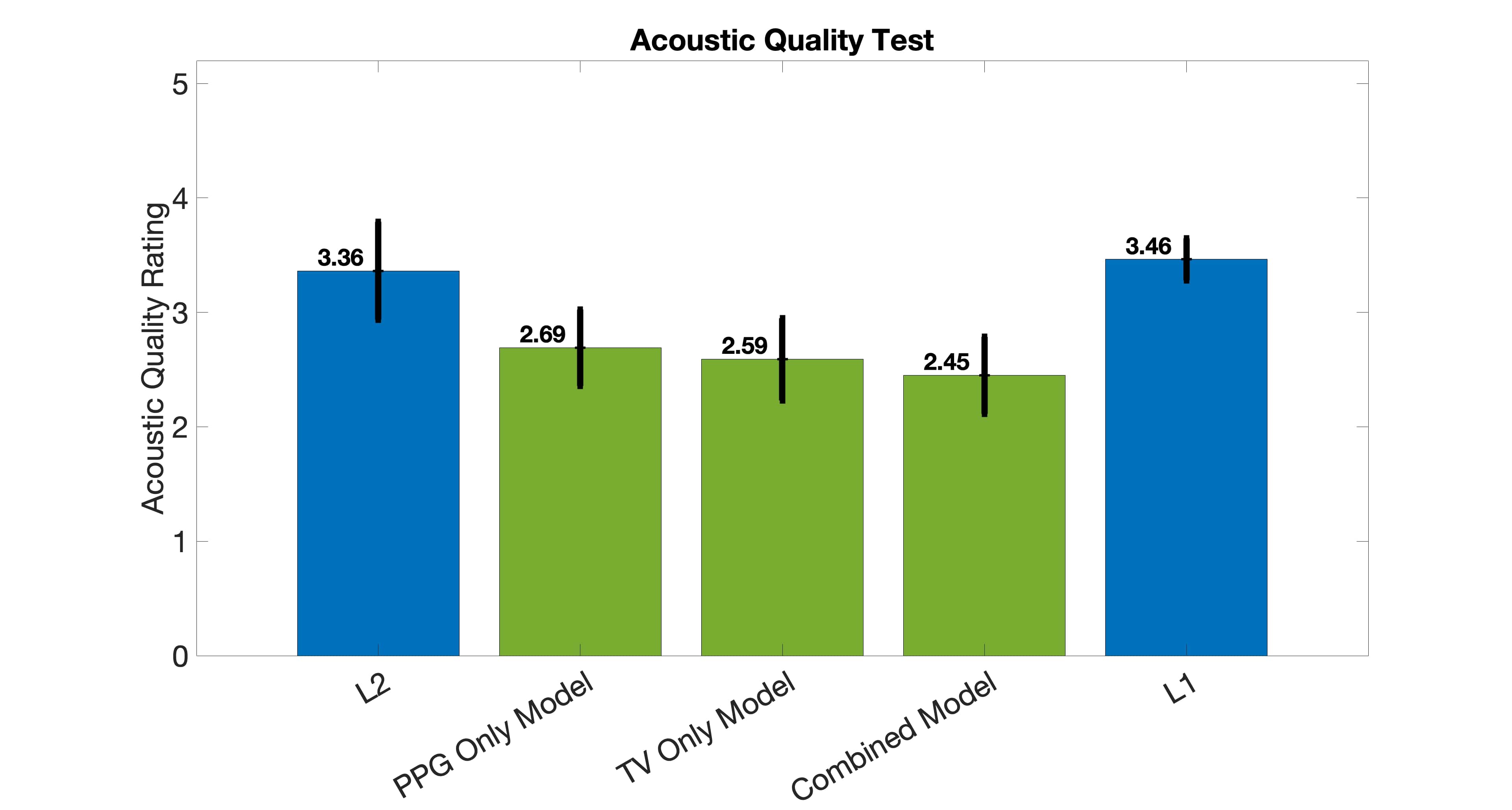}
  \caption{Bars are MOS mean for each system. References are shaded blue and test systems are shaded green. Error bars are 95\% confidence interval}
  \label{fig:mos}
  \vspace*{-14pt}
\end{figure}

Subjects ranked the acoustic quality of 75 utterances using a 5-point mean opinion score (MOS), where higher scores represented improved acoustic quality while lower scores represented poorer quality (5 - Excellent, 4 - Good, 3 - Fair, 2 - Poor, 1 - Bad). Each subject ranked 15 utterances per test system and an additional 15 utterances per reference system. Two-sample t-tests were performed between all three test systems and both reference systems respectively. A Bonferroni adjusted significance level ($\alpha = 0.005$) was applied to all tests. The combined model was the only system to score significantly lower than the L2 reference ($p < 0.005$). Both the PPG only and TV only systems scored lower than the L2 reference ($p = 0.01, p = 0.006$ respectively) however this difference did not reach statistical significance. All three FAC systems scored significantly lower than the L1 reference. 
Additionally, two-sample t-tests were used to evaluate differences between the three test systems. The PPG only method, TV only method, and combined method performed comparably meaning there was not a statistically significant difference in MOS score between the three systems under test. Results are shown in table \ref{tab:test1and2_results} and figure \ref{fig:mos}.

\vspace*{-6pt}
\section{Discussion and Future Work}
\label{sec:duscussion}
\vspace*{-3pt}

This work proposes a multi-task learning based acoustic model architecture to incorporate TVs from a speech inversion system, with conventionally used PPGs, to extract a compact speech representation (BNFs) for the downstream task of accent conversion. Three acoustic model variants (PPG only, TV only and Combined) were developed to extract BNFs and then used for training a speaker conditioned seq2seq model architecture in unsupervised fashion. The pre-trained seq2seq model is then used to perform accent conversion on a held out unseen speaker set in zero-shot fashion. The resulting accent converted samples were evaluated based on objective and subjective measures to understand the feasibility of incorporating articulatory features in foreign accent conversion, and the overall performance of the implementation.

The MOS scores from the Acoustic Quality Test and the MCD scores from the objective evaluations suggest that the PPG only acoustic model variant has slightly better synthesis quality in accent converted samples. However, the WER scores and the Accent Conversion Test scores suggest that the `TV only' and `Combined' acoustic model variants are better at removing the foreign accent compared to the PPG only variant. Additionally, the t-SNE analysis reveals that the speaker identity is better preserved in the TV-only model, which is a crucial aspect of accent conversion. Overall, these findings confirm the feasibility in incorporating articulatory features in accent conversion tasks. Moreover, they highlight the potential enhancement in accent conversion performance by integrating articulatory representations obtained from an acoustic-to-articulatory speech inversion system with a novel multi-task learning-based acoustic model. 

It is important to note that the accent conversion pipeline used in this work requires native reference speech at the time of inference. Even though this limits the use of the proposed system in real-world applications, this work proposes an important direction towards improving the current accent conversion pipelines by incorporating articulatory representations. Moreover, the authors plan to circumvent the need for native reference speech by training a separate translator model (another seq2seq model) in the future. Here the translator model will learn a mapping from non-native BNFs generated from the acoustic model to corresponding native speech's BNFs generated from the same acoustic model. The pre-trained translator model can then be used to generate corresponding native speech's BNFs (left branch of Figure \ref{fig:fac_architecture} in Accent conversion stage) to replace the need for reference native speech. 



     
     

\bibliographystyle{IEEEtran}
\bibliography{mybib}

\end{document}